# OOG- Optuna Optimized GAN Sampling Technique for Tabular Imbalanced Malware Data

S.M Towhidul Islam Tonmoy
Department of Electrical and Electronic Engineering
Islamic University of Technology
Gazipur-1704, Bangladesh
towhidulislam@iut-dhaka.edu

S.M Mehedi Zaman
Department of Electrical and Electronic Engineering
Islamic University of Technology
Gazipur-1704, Bangladesh
mehedizaman@iut-dhaka.edu

*Abstract*— Cyberspace occupies a large portion of people's life in the age of modern technology, and while there are those who utilize it for good, there are also those who do not. Malware is an application whose construction was not motivated by a benign goal and it can harm, steal, or even alter personal information and secure applications and software. Thus, there are numerous techniques to avoid malware, one of which is to develop samples of malware so that the system can be updated with the growing number of malwares, allowing it to recognize when malwares attempt to enter. The Generative Adversarial Network (GAN) sampling technique has been used in this study to generate new malware samples. GANs have multiple variants, and in order to determine which variant is optimal for a given dataset sample, their parameters must be modified. This study employs Optuna, an autonomous hyperparameter tuning algorithm, to determine the optimal settings for the dataset under consideration. In this study, the architecture of the Optuna Optimized GAN (OOG) method is shown, along with scores of 98.06%, 99.00%, 97.23%, and 98.04% for accuracy, precision, recall and f1 score respectively. After tweaking the hyperparameters of five supervised boosting algorithms, XGBoost, LightGBM, CatBoost, Extra Trees Classifier, and Gradient Boosting Classifier, the methodology of this paper additionally employs the weighted ensemble technique to acquire this result. In addition to comparing existing efforts in this domain, the study demonstrates how promising GAN is in comparison to other sampling techniques such as SMOTE.

## I. INTRODUCTION

Internet has had a significant impact on moulding modern society from a global perspective, which is seen in every aspect of people's lives. Everyone, from infants to the elderly, relies on it daily from early morning until night time [1]. However, when people become increasingly dependent on something, it is possible that harmful activities would increase in its vicinity. In the case of digital computers and electronics, this harmful behaviour might manifest in the form of Malware. In other words, malware is software that interferes with the functions of other software, preventing the original software from operating normally [2]. While this may appear to be relatively harmless to the general public, it has shown to be quite the reverse. Malware infects computers and mobile devices on a daily basis with the intent of stealing personal information such as address, bank credentials, and contact information, hijacking personal files and demanding ransom to regain access, and attacking ATMs and bank servers to cause significant monetary and infrastructure damage [3]. Therefore, it has become imperative to prevent these cyberattacks by implementing countermeasures as the reports of malware attacks show 5.4 billion attacks in 2021 and continue to rise.

Antivirus software are programs designed to safeguard computers and mobile devices against malware by identifying and eliminating it. These tools use a database to maintain track of existing malware source codes so that when malware enters a system, it may be promptly recognised by comparing its characteristics to those of existing malware samples [4]. However, this strategy is not unknown to attackers, who employ a variety of techniques to infect personal computers without being detected by antivirus software. Polymorphism is such a behaviour in which malevolent entities constantly change and adapt to evade detection by security technologies. For example, more than 12,000 variations of WannaCry malware have been identified, which was one of the worst cyberattacks in recent history [5]. To detect all of these malware variations and potentially remove them from the system would require a large number of malware samples with slightly altered source code to elude antivirus software.

Generative Adversarial Networks (GAN) is a sampling approach that estimates the likely distribution of sample data and produces fresh samples from the distribution. It has been mostly utilised in the fields of image computing, natural language processing, and object detection; nevertheless, due to its adaptable capacity to create new samples, it is currently now utilised in the field of cybersecurity [6]. Malware variants can be exhaustively detected and categorised as positive or negative using GAN samples. In addition, by the incorporation of Machine Learning (ML) techniques, the identification of positive malware samples can be anticipated and validated with a specific quantity of data. In this study, the synthesis of malware samples using GAN and subsequent validation of those samples using genuine malware samples was performed using a simplified binary classification approach. This was done for the IEEE Big Data Cup Challenge 2022 on "GAN-Based Training for Binary Classifier" concentrating on track-2 which is the numerical raw data-based challenge.

The remainder of this paper will discuss the subsequent literature on malware detection techniques as well as the prevalence of GAN sampling over other conventional sampling techniques, such as SMOTE, with a focus on binary classification, the methodology used for the competition, the results of the study in terms of prediction metrics, and the future prospects concluding the research.

## II. LITERATURE REVIEW

In recent years, numerous studies have been conducted on malware detection and the sampling approaches utilized in the event of imbalanced data. Pajouh et al. focused on the identification of malicious software on Mac OS X using a variety of standard ML methods, with Decision Tree (DT) exhibiting the greatest accuracy of 96.62 % utilising SMOTE as the sampling approach. Other prediction measures, such as F1 score, precision, and recall, were not evaluated [7]. Li et al. discussed the detection of Android malware using a deep learning (DL) model in a separate paper. This model was comprised of a three-layered structure with 1024 neurons per layer. Instead of focusing on accuracy, they calculated F1 score, precision, and recall to be 96.08%, 97.16%, and 95.22%, respectively. Since the dataset was not imbalanced, there was no need for a sampling procedure [8]. Onwuzurike et al. attained an F1 score of 92% using their static and dynamic techniques to malware classification. The model employed here is unique, as are the stimulators for the models [9]. Similar to our research, Sharma and Gupta used the Androzoo dataset to get an accuracy, F1 score, precision, and recall of 92.98%, 92.25 %, 91.53%, and 92.98%, respectively. However, there is no mention of the sample technique used, despite the fact that the dataset was significantly imbalanced [10]. In their study on malware detection via fuzzy sampling, Khoda et al. presented a novel application of unsupervised ML approaches in this field. They achieved 99.49% accuracy with the highest F1 score of 96.20%. The sampling strategy was the SMOTE algorithm. Notably, in the case of malware detection samples, the F1 score is more important than the accuracy [11, 12]. The authors of a recent work by Talbi et al. explored the significance of characteristics in malware detection and developed a DL model to evaluate their findings. In addition, they worked on the Androzoo dataset, where they attained an F1 score of 98.28% and an accuracy of 97.70%. Their prediction metrics were highly promising, despite the absence of any sampling approaches [13].

This work investigates the potential use of improved sampling strategies, with an emphasis on the use of GAN. Although the difference in prediction metrics may not indicate a major improvement, this approach is nonetheless marginally superior and may yield promising results when applied to other malware datasets. In addition, the tuning approach for GAN parameters utilizing an automatic hypertuning algorithm such as Optuna is presented in the following sections. Furthermore, the research describes a unique architecture based on the GAN and Optuna algorithm for detecting malware samples.

## III. METHODOLOGY

This section of the paper focuses on the GAN-based data analysis techniques, from data gathering to result extraction. It is separated into multiple sections to explain the various approaches taken. Figure 1 depicts the entirety of the workflow of this study.

### A. Dataset Collection

The dataset was extracted from the Androzoo database, which contained various VirusTotal-validated malware as well as a number of safe applications from trusted sources such as the Google Play Store and the smartphone's operating system [14]. The dataset comprises of fixed-time-window samples displaying system call monograms. In the dataset which is a comma separated values (.csv) file, there are 128 columns of input features and 1 column of output class titled 'Label'. 4465 examples of malware apps include categorical values in the 'Label' column of the target predictor, where '0' indicates safe apps and '1' indicates malicious apps.

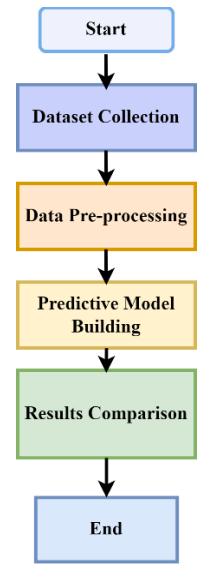

Fig. 1. Workflow diagram.

### B. Data Pre-processing

Initially, the dataset was examined for missing values, but fortunately none were found. Therefore, the pre-processing began with a correlation analysis between the features.

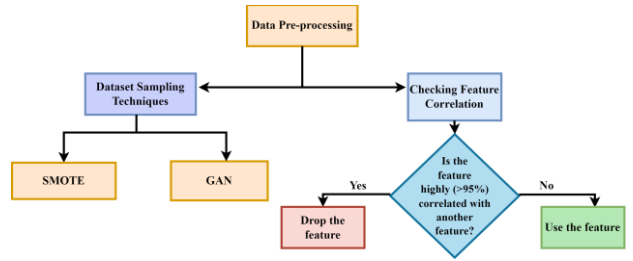

Fig. 2. Data pre-processing flowchart.

#### 1) Checking Correlation

The correlation between all 128 feature columns was examined first. If the correlation between two columns was larger than 95%, one of the columns was dropped. Using this procedure, 28 input feature columns were removed, bringing the total number of columns to 100. This procedure was also proven by its ability to prevent multicollinearity [15].

#### 2) Dataset Sampling Techniques

##### a) SMOTE

The target attribute 'Label' had an uneven distribution of its 0s and 1s. In the imbalanced set, the number of 0s was 3000 and the number of 1s was 1465. The result of training an ML system with this unbalanced dataset and distorted class distribution is skewed towards the majority class. Undersampling or oversampling, which results in class-balanced data, can solve the issue. Synthetic Minority Oversampling Technique, often known as SMOTE, is a frequently used oversampling technique to reduce imbalance in a dataset, and it was also employed in this study [16]. Consequently, the dataset had an equal number of class samples after sampling (3000 each).

##### b) GAN

GAN is another sampling technique where it is capable of mimicking a particular data distribution. GANs are comprised of two neural networks, one trained to produce data and the other taught to differentiate between fake and real data (thus the adversarial nature of the model). Though it's more prevalent in the image/video data, GAN is being used in raw-numerical data as well. There are different versions of GANs and CTGAN, TVAE, and the Copula GAN models were utilized to create malware data in this study. There are numerous parameters for these GAN models, including epochs, generator and discriminator dimensions, embedding dimensions, etc., which can be hyper-tuned to generate superior samples [17-23]. Thus, the hyperparameter tuning algorithm, Optuna, which has been used in this study, will be discussed in the next paragraph.

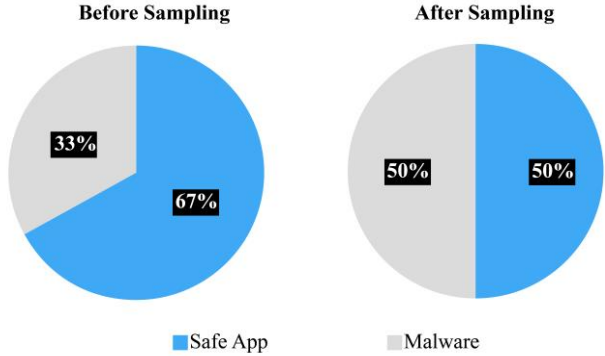

Fig. 3. Percentage of Safe App and Malware before and after sampling.

Optuna is a framework for autonomous hyperparameter optimization software created specifically for machine learning. It has an imperative, define-by-run user interface. Thanks to the define-by-run API, the code built with Optuna is very modular, and the search spaces for hyperparameters may be constructed dynamically [24]. For tweaking the GAN model in this study, the optimal hyper parameters were determined using optuna. The optimal number of epochs, generator learning rate, discriminator learning rate etc. were also evaluated using this method. The Copula GAN model produced the most effective malware samples considering the tuned hyperparameters. The optimal values recovered from the abovementioned method are tabulated in table 1:

TABLE I. OPTIMAL VALUES OF THE GAN MODEL

| Parameter | Value |
|---|---|
| Epochs | 200 |
| Generator dimension | (32, 288) |
| Discriminator dimension | (224, 192) |
| Embedding dimension | 416 |
| Generator Learning Rate | 1.091e-3 |
| Discriminator Learning Rate | 5.402e-3 |

Following the sampling phase, the development of predictive models is the next stage forward.

### C. Predictive Model Building

#### 1) Baseline Models

Five supervised models namely CatBoost, XGBoost (XGB), LightGBM (LGBM), Extra Trees Classifier (ETC) and Gradient Boosting Classifier (GBC) were implemented at first. These boosting classifiers were chosen because they typically perform better in case of binary classification of the target class. These algorithms can also handle nonlinear relationships between features quite effortlessly [25].

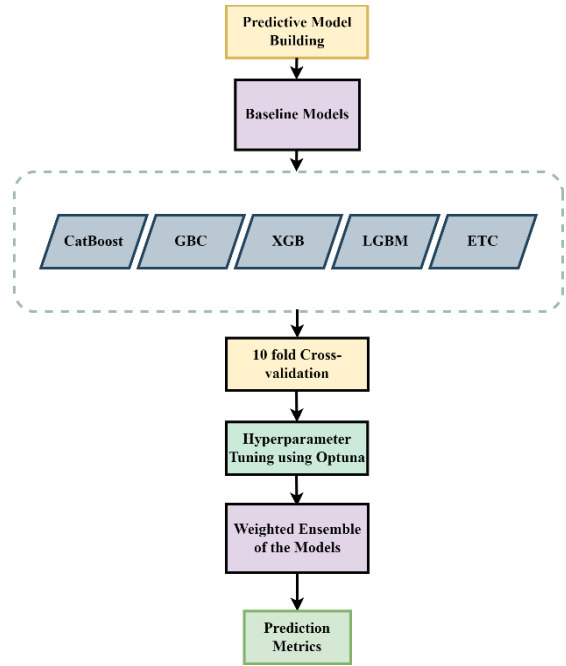

Fig. 4. Predictive model building flowchart.

#### 2) Cross-Validation

Cross-validation (CV) statistically assesses and compares algorithms by separating data into two segments: one for training a model, and the other for verifying the model once it has been trained. In conventional CVs, the training and validated sets must cross over in successive rounds for every data point to be validated [26]. K-fold CV, an approach that divides a dataset into K sets of roughly similar size that are then used to develop the model while the left-out samples are used to test it, is one of the most prominent CV methods. During each iteration of this approach, each of the K folds is assigned as validation data. The most typical range for K is between 5 and 10 [27]. The baseline models in this study were cross-validated into 10 folds.

#### 3) Hyperparameter Tuning

To achieve the best results, ML models must be customised with parameters pertinent to the datasets being used. However, determining the proper parameters is a difficult task that could take a substantial amount of time if the dataset is enormous. Consequently, several ways have been devised to automatically extract these customised parameters. Optuna is one such technique where the pruning integration of the algorithms was utilised to eliminate ineffective trials and accelerate the testing of the models [28]. Using early stopping rounds, the models were prevented from over-fitting. In order to improve the performance of these models, additionally, the significance of their features was evaluated [29]. Table 2 illustrates the best hyperparameters of the algorithms used in the research.

TABLE II. BEST HYPER PARAMETERS OF THE ALGORITHMS

| Model | Hyperparameters |
|---|---|
| XGB | alpha=3.627 |
|  | random_state=101 |
|  | learning_rate= 0.04 |
|  | lambda=0.0015 |
|  | colsample_bytree=0.462 |
|  | subsample=0.598 |
|  | n_estimators=1125 |
|  | max_depth=22 |

| | | |
|---|---|---|
| LGBM | | min_child_weight=2.057 |
| | | learning_rate=0.136 |
| | | n_estimators=10049 |
| | | random_state=101 |
| | | reg_alpha=8.75 |
| | | reg_lambda=2.179e-10 |
| | | colsample_bytree=0.8 |
| | | subsample= 0.624 |
| | | max_depth=20 |
| | | min_child_samples=50 |
| | | num_leaves=232 |
| ETC | | random_state=101 |
| | | n_estimators=950 |
| | | min_samples_split=2 |
| CatBoost | | random_state=101 |
| | | colsample_bylevel=0.093 |
| | | depth=11 |
| | | boosting type='Plain' |
| | | bootstrap_type='MVS' |
| GBC | | learning_rate=0.211 |
| | | n_estimators=158 |
| | | random_state=101 |
| | | max_depth=5 |
| | | subsample=0.8 |

*4) Weighted Ensembling with Hyper-Tuned Classifier*

Ensemble learning is the concept of combining multiple classifiers to provide predictions that are frequently more precise than those of individual classifiers. In addition, it emphasises the robustness and stability of the prediction models in order to improve the metrics to a higher level. Weighted average ensembles presume that certain models in the ensemble are more competent than others and give them a larger contribution to the prediction process. The weighted average or weighted sum ensemble, which is used in this study, is an extension of voting ensembles, which assume that all models are equally skilled and contribute proportionally to the ensemble's predictions [30]. Gathering all of the best hyperparameters, the hypertuned algorithms were ensembled together with the help of a voting classifier. Both the 'soft' and 'hard' voting strategies were implemented and the former one provided the better results. The weight of the models was derived using the different models' f1 weighted scores [31, 32]. Thus, model construction came to an end.

IV. RESULTS AND DISCUSSIONS

This study's predictive model was used to evaluate the imbalanced dataset before comparing it to the balanced datasets using SMOTE and GAN. The comparison is made with the average scores for Accuracy, Precision, Recall, and f1 in the following table 3.

TABLE III. COMPARISON OF MODELS AMONG IMBALANCED, BALANCED (SMOTE) AND BALANCED (GAN) DATASETS

| Model | Accuracy (%) | | | Precision (%) | | | Recall (%) | | | F1 score (%) | | |
|---|---|---|---|---|---|---|---|---|---|---|---|---|
| | Imbalanced | Balanced (SMOTE) | Balanced (GAN) | Imbalanced | Balanced (SMOTE) | Balanced (GAN) | Imbalanced | Balanced (SMOTE) | Balanced (GAN) | Imbalanced | Balanced (SMOTE) | Balanced (GAN) |
| XGB | 96.90 | 97.20 | **97.28** | 97.97 | 96.99 | **98.50** | 92.49 | 94.40 | **96.03** | 95.15 | 95.67 | **97.24** |
| LGBM | 97.44 | 96.43 | **97.81** | 97.95 | 95.36 | **98.38** | 94.19 | 93.72 | **97.23** | 96.03 | 94.52 | **97.80** |
| ETC | 96.93 | 96.93 | **97.63** | 97.36 | 96.24 | **98.11** | 93.17 | 94.33 | **97.13** | 95.22 | 95.27 | **97.62** |
| CatBoost | 97.33 | 97.11 | **97.76** | 98.42 | 97.11 | **99.00** | 93.37 | 93.99 | **96.50** | 95.83 | 95.52 | **97.73** |
| GBC | 97.06 | 97.02 | **97.60** | 97.72 | 96.58 | **98.53** | 93.24 | 94.26 | **96.63** | 95.42 | 95.40 | **97.57** |
| Weighted Ensembled | 97.53 | 97.24 | **98.06** | 98.63 | 97.26 | **98.98** | 93.78 | 94.26 | **97.13** | 96.15 | 95.73 | **98.04** |

Therefore, from the table 3, it is quite clear that the balanced dataset with GAN sampling outperformed the rest of the 2 datasets in all four-prediction metrics.

The balanced dataset with GAN sampling showed the best prediction metrics among the three datasets. It has been depicted in the following Fig. 5.

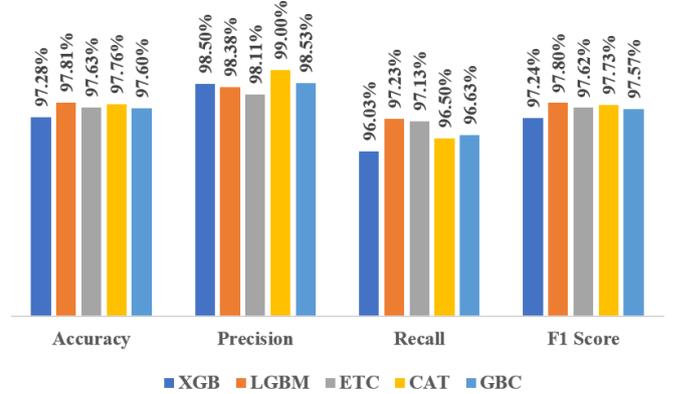

Fig. 5. Comparison of the models' performance with balanced dataset (GAN).

To illustrate the weighted ensembled prediction metrics for all the models of all three datasets, Fig 6. can be portrayed.

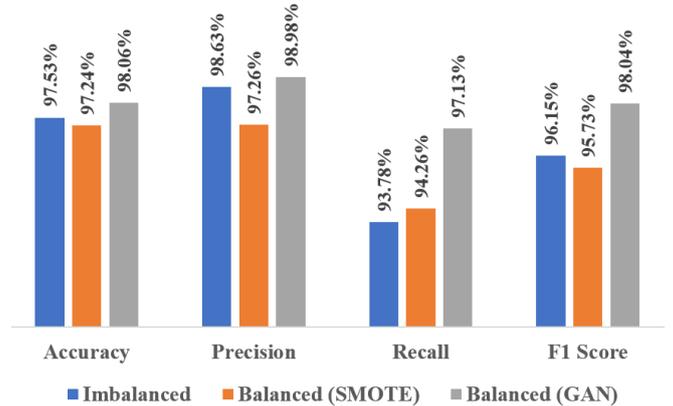

Fig. 6. Comparison of weighted ensemble models' performance of all three datasets.

The Root Mean Squared Error (RMSE) is another prediction metric that square roots the mean of the square of all the errors [33]. The RMSEs of each of the three datasets have been compared in Table 4.

TABLE IV. COMPARISION OF RMSES OF THE MODELS WITH IMBALANCED, BALANCED (SMOTE) AND BALANCED (GAN) DATASETS

| Model | RMSE | | |
|---|---|---|---|
| | Imbalanced | Balanced (SMOTE) | Balanced (GAN) |
| XGB | 0.175 | 0.166 | **0.164** |
| LGBM | 0.159 | 0.187 | **0.147** |
| ETC | 0.175 | 0.174 | **0.153** |
| CatBoost | 0.163 | 0.169 | **0.149** |
| GBC | 0.170 | 0.172 | **0.154** |
| Weighted Ensembled | 0.156 | 0.165 | **0.138** |

According to the aforementioned table, the dataset that was balanced using GAN has the lowest error in terms of RMSE.

Now, the best results from the study's performance can be compared with the previous literature in this domain of malware detection. Table 5 can be shown to see the comparisons.

TABLE V. COMPARISON OF PERFORMANCE WITH PREVIOUS WORKS.

| References | Sampling Technique | Algorithm | Accuracy (%) | Precision (%) | Recall (%) | F1 Score (%) |
|---|---|---|---|---|---|---|
| [7] | SMOTE | DT | 96.62 | - | - | - |
| [8] | - | DL | - | 97.16 | 95.22 | 96.08 |
| [9] | - | Static and Dynamic | - | - | - | 92.00 |
| [10] | - | - | 92.98 | 91.53 | 92.98 | 92.25 |
| [11] | SMOTE | Unsupervised | 99.49 | - | - | 96.20 |
| [13] | - | DL | 97.70 | - | - | 98.28 |
| **This study** | **GAN** | **Weighted Ensembled Classifier** | **98.06** | **99.00** | **97.23** | **98.04** |

This study demonstrates that malware samples collected using GAN are superior than those collected using other sampling techniques, such as SMOTE. In addition, the ML architecture of adjusting the GAN using Optuna is relatively new to this field and has been found to be highly beneficial. The Androzoo database is just one of the datasets used to classify malware and safe applications; there are others that also demand enough attention as cyberspace becomes increasingly populated with malware upgrades. Consequently, the use of GAN in these instances could be advantageous for a secure internet browsing experience and possibly a malware-free digital environment.

## V. CONCLUSION

As the internet and digital technologies such as smartphones have grown ubiquitous in today's society, it is crucial to monitor and safeguard the general public from malicious programmes and malware that may illegally take, change, or delete one's personal data. However, as time passes, both the quantity and severity of malware increase at an alarming rate. Therefore, it is the responsibility of security frameworks and software to prevent these malwares from accessing the system and, if they do enter, to quickly detect and delete them. To accomplish this, security apps must be continually updated with new malware samples so they can compare and determine if an app is malicious or not. Existing security methods include sampling techniques such as SMOTE, GAN, etc. to generate more synthetic samples that resemble actual malware in order to keep up with the daily updates.

This paper emphasises GAN sampling over SMOTE, and the results support this position. Additionally, the addition of Optuna for modifying GAN parameters has marginally improved the results, which could be an important element in future malware detection studies. The acquired data also demonstrate a high F1 measure of 98.04 %. In the future, the GAN sampling technique may outperform other contemporary sampling strategies even for classifications based on raw data, despite the fact that GAN is predominantly utilised for image-based datasets.